\documentclass[letterpaper,11pt]{article}
\RequirePackage[OT1]{fontenc}
\RequirePackage{amsthm,amsmath}
\RequirePackage[numbers]{natbib}
\RequirePackage[colorlinks,citecolor=blue,urlcolor=blue]{hyperref} 
\usepackage[numbers]{natbib}
\usepackage{amssymb}
 \usepackage{amsthm}
 \usepackage{amsmath}
\usepackage{amsfonts}
\usepackage[english]{babel}
\usepackage[utf8]{inputenc}
\usepackage{hyperref}
\usepackage{bm}
\usepackage{float}

\begin{document}
\title{New subluminal and superluminal spacetime measures}
\author{Benjam\'{i}n Calvo-Mozo
\footnote{{\em bcalvom@unal.edu.co}}\\
\footnotesize{Universidad Nacional de Colombia, Observatorio Astron\'omico Nacional, Bogot\'a, Colombia}}
\date{}
\maketitle
\begin{abstract}
In present work the author presents a new set of spacetime measures for both, subluminal and superluminal motion regimes, which do not diverge for the speed of light in the former case and each regime has its own light-like cone which warrants causality to each regime. For events located in the vicinity of the light cone and inside it, there exists the possibility to transit across the speed of light value in a discrete way, in terms of the very small constant which we introduce here to prevent the divergence of our spacetime measures at the speed of light.
\end{abstract}
\section{Introduction}
It is a real challenge for humankind to think of a serious way of taking a body at rest, accelerates it to near the speed of light, and then crossing safely the {\it c-barrier}, that is, carrying it to so-called warp speeds ($v>c$). This was one of the goals of NASA's Breakthrough Propulsion Physics Program \citep{Millis}.  Present day physics does not allow it, for there is a theoretical impediment given by the special theory of relativity, for it requires an infinite energy consumption because speeds vary in a continuous manner. \citet{Alcubierre} found a solution of Einstein's field equations, where spacetime contracts in front of some spacecraft and expands behind it, so that for an asymptotic observer it seems like moving with a superluminous speed, though the spacecraft really moves with a speed lesser than the speed of light in vacuum; this solution requires negative mass (exotic) matter.\\

Nowadays, the spacetime arena suitable for subluminal motions uses Lorentz transformations to compare positions and time measures of material bodies as observed by two inertial frames in uniform relative motion. From these equations we obtain the well known Einstein's rule of velocities addition. This rule implies that the speed of light in vacuum is both, an invariant for all inertial frames and also the maximum allowable speed. The equations for this rule of velocities addition do not diverge when $v=c$, where $v$ is the uniform relative speed between two inertial frames, whilst Lorentz transformations diverge in the expressions for $x$ and $t$ at the speed of light ($v=c$). \\

Here we propose a generalization of Lorentz transformations which do not diverge for $v=c$, and which also satisfy the mentioned rule of velocities addition. We do that by adding a positive very tiny constant, say $\epsilon^2$, into the square root which appears in Lorentz transformations, such that it prevents the divergence of spacetime measures at the speed of light; as a second step, we also modify the other two equations ($y,z$) in order to keep the invariance of light speed in vacuum. However, our new spacetime measures approximate to Lorentz transformations for most relativistic speeds, but they differ drastically when $\beta_1^2=1-v^2/c^2$ is comparable with the constant $\epsilon^2$.  With our new spacetime measures we can think then of taking a body at rest, accelerates it to some speed near but lesser than the speed of light, and make a {\it discrete jump} in $v^2/c^2$ across the unit in terms of $\epsilon^2$, so that the squared dimensionless speed of the material body passes from some value near but lesser than the unit to another one a bit greater than the unit. That is, the transition from subluminal to superluminal motion, and vice versa, could be done in a discrete way. It is necessary also to have adecuate spacetime measures for the superluminal motions regime. We find them in a relatively simple way in present work, stating that if the subluminal regime of motions have an upper and invariant speed $c_1=c$, there could exists another upper and invariant speed $c_2$ valid for the superluminal regime, for we want to preserve causality. If we call $\gamma_1$ the subluminal nondivergent term which contains the square root mentioned above, it takes the value $\epsilon^{-1}$ at the speed of light. We calculate the speed $c_2$ by taking $\gamma_2=2\epsilon^{-1}$ at the speed $c_2$; its value results to be very much greater than $c$: $c_2\simeq\epsilon^{-1}c$.\\

Our spacetime measures suitable for superluminal motions should take the same algebraic form than those for subluminal motions, because we impose the condition that the speed $c_2$ is also a top and invariant speed, this time in the superluminal regime. We write down then spacetime equations for both motion regimes as a single one, distinguishing them only by a subindex $i=1,2$ where the subindex $i=1$ stands for the subluminal motions regime. From these equations we find also the composition rule for velocities addition in a compact and dimensionless expression with the subindex $i=1,2$ specifying again the motions regime.\\

An interesting result of our way of constructing nondivergent spacetime measures is that if we allow that the index $i$ denoting the motions regime can take any integer value $i\geq1$, such that the corresponding $\gamma_i=i\epsilon^{-1}$ at the top and invariant speed $c_i$ of the corresponding speeds interval, we find then the expressions for $\gamma_i$, $c_i$ in the superluminal regimes. Our new spacetime measures also involve another factor $\beta_i=(1-v^2/c_i^2)^{1/2}$ in the expressions of $y,z$. Let us call to $\gamma_i$ and $\beta_i$ measuring factors. We have that the measuring factor $\gamma_i$ takes the maximum value at the top speed of the speeds interval, while $\beta_i$ vanishes at that top speed. The author uses this fact to propose the existence of {\it resonant states} of matter, which determine the description of material bodies depending on their speeds with respect to those of such resonant states. Speed values take non negative real numbers, in such a way that their values are partitioned in intervals, which we can call speed ranges, being the first one closed on both ends, $0\leq v\leq c$, associated to subluminal motions, and the others are open in the lower end and closed in the upper end, $c_{i-1}<v\leq c_i$ for any integer $i\geq2$. These speed intervals or ranges correspond to superluminal motions. The author thinks that this partition of speeds in intervals where the top speed of each interval follows a generalized Einstein rule for velocities addition, warrants the preservation of causality at all motion regimes, the subluminal one and all superluminals. 
\section{Subluminal motion}
In present work, we want to find spacetime measures which preserve the invariance of light speed in vacuum and which do not diverge for light speed. The last condition does not imply the existence of relative motions between inertial frames at that top speed, because from any inertial frame light speed takes the same value due to its invariance. The choice of no divergence at light speed is to assure that when bodies speeds approach extremely near the speed of light, their squared dimensionless speeds can vary in terms of a very tiny constant, that is, in a discrete way. Let $\epsilon^2$ be a real positive dimensionless constant, very small with respect to the unit; later in this paper we will estimate its value as of the order of $10^{-54}$ -cf. eq.(\ref{epsilon2}). Let us add this constant within the square root which appears in Lorentz transformations, so that our new $x$ and $t$ measures are:
\begin{equation}\label{Leps-1}
x=\frac{x'+uct'}{\sqrt{1+\epsilon^2-u^2}},\quad ct=\frac{ct'+ux'}{\sqrt{1+\epsilon^2-u^2}}
\end{equation}
where $u=v/c$. To preserve light speed in vacuum as the maximum and invariant speed for all inertial frames, we ought to modify the expressions for $y$ and $z$ as well:
\begin{equation}\label{Leps-2}
y=\left(\frac{1-u^2}{1+\epsilon^2-u^2}\right)^{1/2} y',\quad z=\left(\frac{1-u^2}{1+\epsilon^2-u^2}\right)^{1/2} z'
\end{equation}
We see that eqs.(\ref{Leps-1}),(\ref{Leps-2}) approach to Lorentz transformations for most relativistic speeds, that is when $1-u^2>>\epsilon^2$ holds. We can rewrite eqs.(\ref{Leps-1}),(\ref{Leps-2}) in an interesting algebraic form, which enables the application to other motion regimes. First, let us call $\gamma_1$ to the non divergent term in these expressions:
\begin{equation}\label{gamma1}
\gamma_1 = (1+\epsilon^2-u^2)^{-1/2},\quad u^2=v^2/c^2
\end{equation}
This gamma factor equals $\epsilon^{-1}$ when $v=c$, which is a large quantity but anyway a finite one. With this gamma factor ($\gamma_1$) our $x$, $t$ measures are:
\begin{equation}\label{ecs_xt1}
x = \gamma_i(x'+u_ic_it'),\quad c_it =\gamma_i(c_it'+u_ix')
\end{equation}
\noindent where $i=1$ for the subluminous regime, for which $c_1=c$, $u_1=v/c_1$, and the $\gamma_1$ factor is given by eq.(\ref{gamma1}). Now, our expressions for the $y$ and $z$ measures in the new algebraic form are:
\begin{equation}\label{ecs_yz1}
y = \beta_i\gamma_i y',\quad z = \beta_i\gamma_i z',\quad \beta_i=(1-u_i^2)^{1/2}
\end{equation}
\noindent where $i=1$ for the subluminal motions regime, that is, for $0\leq v\leq c$. We see that for $v=c$, our set of eqs.(\ref{ecs_yz1}) imply the vanishing of $y,z$ measures, while eqs.(\ref{ecs_xt1}) give,
\begin{equation}\label{luz_1}
x = ct= \epsilon^{-1}(x'+ct')
\end{equation}
That is, for our spacetime measures, the light case behaves like a problem with only one degree of freedom. From eqs.(\ref{ecs_xt1}),(\ref{ecs_yz1}) we can obtain Einstein's rule for velocities addition:
\begin{equation}\label{speeds-sum}
U^{2}_{i}=1-\frac{(1-U'^{2}_{i})(1-u^{2}_{i})}{(1+{\bf u_{i}}\cdot{\bf U'_{i}})^{2}}
\end{equation}
\noindent where $\mathbf{u_i}\cdot\mathbf{U'_i}$ stands for a dot (scalar) product between 3-vectors of dimensionless velocities, ${\mathbf u_i}=\mathbf{v}/c_i$, $\mathbf{U'_i}=\mathbf{V'}/c_i$, and the Cartesian components of vector $\mathbf{V'}$ are $dx'/dt'$, $dy'/dt'$, $dz'/dt'$; the derivatives of the respective unprimed variables leads to the dimensionless speed $U_i$. Vector $\mathbf{v}$, can be considered as usual, that is, as the uniform relative velocity between two inertial observers, except when its magnitude equals that of light in vacuum, $c$, which we interpret later in present work.\\

The set of eqs.(\ref{ecs_xt1}), (\ref{ecs_yz1}) for space and time measures, can be written in a matrix form and from it we can derive their inverses. In effect, if L is the matrix associated to the complete set of these equations, in a matrix form they are $X=LX'$, where $X$,$X'$ are column vectors with $X=(c_it,x,y,z)^T$, in which $T$ denotes the transpose operation, and $X'$ stands for the respective primed variables column vector. We can easily check that the determinant of the $4\times4$ matrix $L$ equals $\beta_i^4\gamma_i^4$, which does not vanish for $v\neq c_i$; then, we can invert matrix $L$ and obtain the inverse of eqs.(\ref{ecs_xt1}), (\ref{ecs_yz1}) by means of $X'=L^{-1}X$. In this way, the inverse spacetime measures are:
\begin{align}\label{inversas1}
c_it'=\beta_i^{-2}\gamma_i^{-1}(c_it-u_ix),\quad &x'=\beta_i^{-2}\gamma_i^{-1}(x-u_ic_it)\\\label{inversas2}
y'=\beta_i^{-1}\gamma_i^{-1}y,\quad &z'=\beta_i^{-1}\gamma_i^{-1}z
\end{align}
Let us observe that in the subluminal motions regime these expressions reduce to the well known ones associated to Lorentz transformations for $\beta_1^2>>\epsilon^2$. In the previous notation, Lorentz transformations are usually written in a matrix form as $X'=\Lambda X$, so that $X=\Lambda' X'$, where $\Lambda'$ is the inverse matrix of $\Lambda$. In this regime and with the strict condition $u^2<1$ the composition of two $L$ matrices does not give a third $L$ matrix as happened with Lorentz matrices $\Lambda$; instead, we have that:
$$
L_AL_B=\left(1+\frac{\epsilon^2}{\beta_A^2}\right)^{-1/2}\left(1+\frac{\epsilon^2}{\beta_B^2}\right)^{-1/2}\Lambda'_C
$$
\noindent where $\Lambda'_C$ is a Lorentz matrix for the dimensionless velocity ${\bf u}_C={\bf u}_A\oplus {\bf u}_B$, which follows an Einstein's sum for velocities addition like that given by eq.(\ref{speeds-sum}) in a compact form. A consequence of this result, is that our set of spacetime measures do not form a group and also that they do not correspond to coordinate transformations, but anyway they can be interpret as results of measures made by rods and clocks as was the original idea of \citet{Einstein1905,Einsteinlibro}. However, when $\beta_1\gg\epsilon^2$, i.e. in the relativistic regime, we can expand the above expression in powers of $\epsilon^2$ and each term of the expansion is proportional to the Lorentz matrix $\Lambda'_C$. Let us note that our set of eqs.(\ref{ecs_xt1}),(\ref{ecs_yz1}) can also be written in a matrix form as $X=\beta_i\gamma_i\Lambda'X'$; then $L=\beta_i\gamma_i\Lambda'$. Thus, the inverse eqs.(\ref{inversas1}),(\ref{inversas2}) can also be expressed as $X'=\beta_i^{-1}\gamma_i^{-1}\Lambda X$, valid only for $u_i\neq1$. From these equations we can obtain a rule for speeds addition similar in form to eq.(\ref{speeds-sum}), interchanging $U_i$ and $U'_i$, and changing the term with the dot product there by $-{\bf u_{i}}\cdot{\bf U_{i}}$. The metric {\bf g} associated to our new spacetime measures equals $\beta_i^2\gamma_i^2\mbox{\boldmath $\eta$}$, where {\boldmath $\eta$} is the Minkowski metric; then, our metric {\bf g} is conformally flat.\\

For subluminal motions, we can calculate lengths of moving bodies in the customary relativistic way, that is, measuring simultaneously their extreme points; thus, with the aid of eqs.(\ref{inversas1}),(\ref{inversas2}) we obtain for lengths parallel to the motion and perpendicular to it, the values $l_{\parallel}=\beta_1^2\gamma_1 l_o$ and $l_{\perp}=\beta_1\gamma_1 l_o$, respectively, where $l_o$ is the rest length; then, volume of moving bodies varies as $\beta_1^4\gamma_1^3$ times the volume at rest. For clocks in motion, time intervals are seen as $\gamma_1\Delta t_o$, where $\Delta t_o$ is the corresponding lapse of proper time. However, all expressions considered here reduce to the respective relativistic ones when $\beta_1^2>>\epsilon^2$.
\section{Superluminal motion}
Let us observe that before reaching the speed of light, the gamma factor $\gamma_1$ given by eq.(\ref{gamma1}) takes the values $(n+1)^{-1/2}\epsilon^{-1}$ for $\beta_1^2=n\epsilon^2$. Thus, we can expect that changes in speed around $u^2=1$ are of the type:
\begin{equation}\label{cambio_u2}
\Delta u^2: 1-n_1\epsilon^2\rightleftarrows1+n_2\epsilon^2
\end{equation}
\noindent where $n_1, n_2$ are positive integers. In author's opinion, the superluminal motions regime must have its own causal ordering of events. To accomplish it, we need another limit and invariant speed, say, $c_2>c$, and a new set of spacetime measures which imply a similar (in algebraic form) composition rule for dimensionless speeds but which takes into account$\ c_2$ instead of $c$. We know that eqs.(\ref{ecs_xt1}),(\ref{ecs_yz1}) infer Einstein's rule for velocities addition given here by eq.(\ref{speeds-sum}). Thus, we need corresponding $\beta_2$ and $\gamma_2$ factors, where the subindex $i=2$ stands for superluminal motions under the $c_2$-cone regime. For the $\beta_2$ expression we have that it must vanish when $v=c_2$. Then, if $u_2=v/c_2$, one has that $\beta_2=(1-u_2^2)^{1/2}$ will satisfy this condition. To find an expression for $\gamma_2$, let us take into account that it comes with an accumulated $\epsilon^{-1}$, and that it goes up as $u^2$ increases, till one reaches the new maximum speed, say, $c_2$. As the gamma factor $\gamma_1=\epsilon^{-1}$ for the speed $c_1=c$, we reasonably assume that $\gamma_2=2\epsilon^{-1}$ when $v=c_2$. We can obtain this result combining positive and negative powers of $\epsilon^2$, such that:
\begin{equation}\label{gamma2}
\gamma_2 =\epsilon^{-1}+(\epsilon^{-2}+1+\epsilon^2-u^2)^{-1/2}
\end{equation}
From this expression we obtain the new limit speed, which results to be very much larger than the speed of light in vacuum:
\begin{equation}\label{c_2}
c_2=(\epsilon^{-2}+1)^{1/2}c\simeq\epsilon^{-1}c \sim5\times10^{26}c
\end{equation}
This is a real huge speed! The $\gamma_1$ factor given by eq.(\ref{gamma1}) is valid for squared dimensionless speeds in the range $0\leq u^2\leq 1$, that is, in the subluminal regime, while the $\gamma_2\,$factor given by eq.(\ref{gamma2}) applies in the range of squared dimensionless superluminal speeds: $1< u^2\leq\epsilon^{-2}+1$; in both cases $u=v/c$. Lengths of moving bodies and time intervals given by clocks in the superluminal regime can be derived using the same procedures as done for the subluminal case: $L_{\parallel}=\beta_2^2 \gamma_2 l_o$, $L_{\perp}=\beta_2\gamma_2l_o$, $\Delta t=\gamma_2\Delta t_o$. These expressions have an interesting behaviour as, for in all of them appears a factor on the order of $\epsilon^{-1}$ for most conceivable speeds. In effect, if $u_2^2\ll1$, then $\beta_2\simeq(1-u_2^2/2)$ and $\gamma_2\simeq(\epsilon^{-1}+\epsilon/\beta_2)$, which in turn can be approximated to the unit and to $\epsilon^{-1}$, respectively. At the second limit speed, $c_2$, eqs.(\ref{ecs_xt1}),(\ref{ecs_yz1}), give $x=c_2t=2\epsilon^{-1}(x'+c_2t')$, which is similar in form to eq.(\ref{luz_1}) except for a factor of 2, whilst measures $y,z$ vanish at the new limit speed $c_2$. That is, it corresponds to a problem with one degree of freedom.
\section{Higher superluminal regimes}
We will find now the appropriate expressions for $\gamma_i$, $c_i$, for all superluminal regimes ($i\geq2$), assuming we allow that the subindex $i$ can take all positive integer values. The one for $\beta_i$ is given by eq.(\ref{ecs_yz1}). We see that eqs.(\ref{ecs_xt1}),(\ref{ecs_yz1}) imply eq.(\ref{speeds-sum}) for speeds addition in a compact form, which let $c_i$ as the maximum speed in the range $c_{i-1}<v\leq c_i$, for any integer $i\geq2$. We obtain $\gamma_i$ for any $i\geq2$, using only positive and negative powers of $\epsilon^2$, considering that it has an accumulated $(i-1)\epsilon^{-1}$ from the previous speeds range and that it should equal to $i\epsilon^{-1}$ at the top speed $c_i$ of the respective speeds interval. Further, in the expression for $\gamma_2$ there are terms with positive and negative powers of $\epsilon^2$, then we shall use higher positive and negative powers of it keeping ``symmetry" in these powers, that is, if there appears the power $\epsilon^{-2k}$ there appears also the power $\epsilon^{2k}$. Using these considerations we have:
\begin{equation}\label{gammai}
\gamma_i =(i-1)\epsilon^{-1}+ \epsilon^{i-2}\left[\sum_{k=1}^{i-1}\left(\epsilon^{-2k}+\epsilon^{2k}\right)+1-u^2\right]^{-1/2}
\end{equation}
The condition imposed on the main measuring factor $\gamma_i$ at the maximum speed of the associated range, gives us the expression of $c_i$ for $i\geq2$:
\begin{equation}\label{luz-i}
(c_i/c)^2 =\sum_{k=1}^{i-1}\epsilon^{-2k} + \sum_{k=0}^{i-2}\epsilon^{2k}
\end{equation}
\noindent which reduces to the expression given by eq.(\ref{c_2}) for $i=2$. For any $i$ eqs.(\ref{c_2}),(\ref{luz-i}) tell us that in first approximation $c_i\simeq\epsilon^{-i+1}c$. Any transition between two speed ranges, say around $(c_i/c)^2$, $i\geq1$, should be done in a discrete way, for it involves two different couple of factors $\beta_i$,$\gamma_i$; thence, the discreteness of the transition around the speed of light value as stipulated by eq.(\ref{cambio_u2}). For all $c_i$ one has that $y=z=0$, and:
\begin{equation}\label{xt-luz-i}
x=c_it=i\epsilon^{-1}(x'+c_it'),\quad \text{and} \quad x'=c_i t'=\frac{\epsilon x}{2i},
\end{equation}
\noindent which is a generalization of eq.(\ref{luz_1}). The first of eqs.(\ref{xt-luz-i}) is obtained directly from eqs.(\ref{ecs_xt1}) making $u_i=1$, that is, for $v=c_i$. For that value eqs.(\ref{ecs_yz1}) give $y=z=0$. The second of eqs.(\ref{xt-luz-i}) is obtained through a limit procedure, making use of l'H\^opital rule of calculus applied to eqs.(\ref{inversas1}).\\

We find then that instead of having only light speed $c_1=c$ and $c_2$ as maximum speeds for their respective ranges, we have the subluminal regime, and an indefinite number of superluminal regimes, each one with a top speed given by eq.(\ref{luz-i}) and corresponding speeds range, $c_{i-1}<v\leq c_i$, with $i\geq2$.
\section{An interpretation}
The appearance of a set of top and invariant speeds, partition speed values in sets of nonnegative real numbers or intervals which we can call speed ranges, being the first one closed in both sides, that pertaining to the subluminal regime, while the others are open in the lower end and closed in the upper value and which correspond to superluminal regimes. This fact and taking also into account that measuring factors $\gamma_i$, $\beta_i$ take the maximum and minimum value, respectively, at the top speed of the respective speeds range, enable us to infer a new property of matter, and is the existence of {\it resonant states} of matter which determine the behaviour of any material body depending on its speed with respect to those of such resonant states.\\
Though our eqs.(\ref{ecs_xt1}),(\ref{ecs_yz1}) do not diverge for $v=c_i$, it does not mean that it serves as a reference frame to describe events, because eq.(\ref{speeds-sum}) derived from these equations imply that anyway it is seen with speed $c_i$ by any (inertial) observer. These resonant states with respect to spacetime measures reduce to only one degree of freedom, for in that case $y,z=0$, and $x=c_it$; further, we have $x'=c_it'$. So we can interpret eq.(\ref{xt-luz-i}). Let us examine the light case. If we take the limit $u\to1$ in eqs.(\ref{inversas1}), consider eq.(\ref{gamma1}) for $\gamma_1$, and the third of eqs.(\ref{ecs_yz1}) for $\beta_1$, one obtains $x'=ct'=\epsilon x/2$. This result enables us to examine the behaviour of our spacetime eqs.(\ref{ecs_xt1}),(\ref{ecs_yz1}) for light, that is, for eq.(\ref{luz_1}). It could tell us that half of the {\it structure} associated to photons lies in $x'$ whilst the other half corresponds to $ct'$. A possible interpretation of this result is that dynamical measures associated to photons reside in every point of that part denoted by $x'$ of photon's structure and measured as $\epsilon^{-1}x'=x/2$; the other half, that pertaining to $ct'$, indicates a tendency to move, that is, to occupy an equal amount of space, just contiguous to the first one, but to be ocuppied a time $\epsilon^{-1}t'=x/2c$ later. As both appear in eq.(\ref{luz_1}) they are integral part of photon's structure. Once the former half occupies the second one, it continues this tendency successively, and maybe this is the reason why it has a wavelike behavior.\\

Going forward in our interpretation of the results presented here, we can think that when photons propagate, or particles move, what are moving are their associated structures defined on the elementary bases of space, understanding by them the finest partition of space. When we go down looking for the finest partition of space, we find lengths on the order of Planck's length, $L_P$. \citet{Planck1899} thought that when we arrive to such lengths, space could be described in a discrete manner, while for \citet{Sakharov1967} lengths of order $L_P$ represent our limits of the concept of space in the sense of localization. Wheeler thought of some kind of pregeometry at such levels (see box 44.5 of \cite{Gravitation}) as the ``basic building" of spacetime. For \citet{Oriti2013}, there should be some kind of {\it atoms} of space; he was based, in part, on the idea of emergence of spacetime proposed by \citet{HW2012}. Taking into account these ideas, the author of present paper proposes here that what we interpret as a point in our spacetime measures given by eqs.(\ref{ecs_xt1}),(\ref{ecs_yz1}), or of their inverses given by eqs.(\ref{inversas1}),(\ref{inversas2}), is something of the order of $L_P^3$ in volume, which we will call here an {\it element} of space or of the structures under discussion, like those associated to photons or particles, or even of some region of space wherein there exists some field. On such elements of structures associated to photons or particles, or of regions of space with fields, we can define geometrical, kinematical or dynamical measures.
\section{Value of $\epsilon^2$}
To estimate a value for $\epsilon^2$ we see first that when $u=0$, eqs.(\ref{Leps-1})-(\ref{gamma1}) give us spacetime measures of type $\delta Y/Y_o\simeq-\epsilon^2/2$ where $\delta Y=Y-Y_o$; in these couple of expressions $Y$ denotes either $x$, $y$, $z$, $ct$, and $Y_o$ stands for either of the primed variables. For the case of our metric we have $\delta g/g_o\simeq-\epsilon^2$ for $u^2<<1$. A possible interpretation is that in any vacuum, even if it is far from the influence of ponderable matter or fields, vacuum has natural fluctuations and then if one put there a test particle it should be influenced by these fluctuations. However, we should look for an observable quantity in order to estimate a value for $\epsilon^2$ by analogy. In author's opinion it is the case of the correction made on the electron spin g-factor, $(g_e-2)/2$, associated to the anomalous magnetic moment of electron, which is on the order \citep{Schwinger,Weinberg} of $\alpha/2\pi$, where $\alpha$ is the fine structure constant, and which results from radiative corrections due to vacuum fluctuations. \citet{Weinberg} employs the concept of ``charge radius" in these calculations as the zone of influence of vacuum fluctuations on the electron; we use here instead the Compton wavelength of electron, using an intuitive simple linear approach. In this way of reasoning, we can assume that electrons and ponderable matter particles can have also structures with some ``internal" dimensionles speed, say, of the type $U_1^{*2}=1-n\epsilon^2$, in the subluminal regime, with some $n$ comparable to the unit, for instance (to put a number, see below) $n=3$. In this case we have $y/y'=z/z'=\sqrt{3}/2$, and:
\begin{equation}\label{x-t-e-1}
x=\frac{1}{2}\epsilon^{-1}(x'+U_1^*ct'),\,\, ct=\frac{1}{2}\epsilon^{-1}(ct'+U_1^*x')
\end{equation}
Then, $x\simeq ct$, for $U_1^*$ almost equals the unit; further, one has -cf. eqs.(\ref{inversas1}): $x'\simeq ct'\simeq\epsilon x$. Thus, according to our previous interpretation for the light case, the kind of structure described by eqs.(\ref{x-t-e-1}) can be seen as having a wavelike behaviour, in first approximation, when it displaces freely. Let $N_o$ be the number of associated electron elements which we take into account for the observed correction on its magnetic moment. We see that the ratio of electron's Compton wavelength over Planck's length gives on the order of $10^{23}$. However, we have not made 3-dimensional considerations; so let us take $N_o$ of the order of a familiar quantity, say, Avogadro's number, but taking it dimensionless: $N_o=N_A\times(1\text{mol})$, where $N_A$ is Avogadro's number. Now,  if the terms within parenthesis are of the type $Y_o$ as discussed above, it equals to $Y-\delta Y$. Following the discussion written above, we can say (intuitively) that $N_o\epsilon$ is observed as $\alpha/2\pi$; then, our very tiny constant equals:
\begin{equation}\label{epsilon2}
\epsilon^2 = \left(N_o^{-1}\frac{\alpha}{2\pi}\right)^2 \simeq 3.7\times10^{-54}
\end{equation}
In our calculations above, we have used $U_1^{*2}=1-3\epsilon^2$ for structures of some ponderable matter particles. This choice is not arbitrary; the reason is that if we can associate some dynamical measure per basic element of spacetime of the type $S(u)=\gamma_i(u)S_o$, where $S_o=\epsilon h$ is a constant, with $h$ the Planck constant, then $S(1)=h$ and $S(U_1^*)=h/2$. The author will explore simple dynamical considerations compatible with our spacetime measures given by eqs.(\ref{ecs_xt1}),(\ref{ecs_yz1}) in another work.
\section{Conclusions}
The first main contribution of present article which the author wants to highlight, is that for the subluminal regime we can construct spacetime measures which, (i) in first approximation give the set of Lorentz transformations, (ii) do not diverge for $v=c$, and (iii) keeps the light speed in vacuum as an invariant for inertial observers under uniform relative motion, thus preserving causality. This set of spacetime {\it measures} are given by eqs.(\ref{ecs_xt1}),(\ref{ecs_yz1}), with $i=1$ for the subliminal regime, or explicitly by eqs.(\ref{Leps-1}),(\ref{Leps-2}). In this regime $c_1=c$, $\gamma_1$ is given by eq.(\ref{gamma1}) and $\beta_1$ by eq.(\ref{ecs_yz1}) with $u_1=v/c_1$. However, we see also that we can take other values of $i$ in eqs.(\ref{ecs_xt1}),(\ref{ecs_yz1}), for instance $i=2$, or even we can consider higher (positive) integer values with appropriate values of $\gamma_i$, $\beta_i$ and $c_i$, which correspond to what we call here superluminal regimes, being the first one that associated to the range of speeds $c<v\leq c_2$, where $\gamma_2$, $c_2$ are given by eqs.(\ref{gamma2}),(\ref{c_2}), respectively; the factor $\beta_2$ is calculated using eq.(\ref{ecs_yz1}) with $u_2=v/c_2$.\\

For arbitrary $i\geq2$, that is, for all superluminal regimes, we have $u_i=v/c_i$, and $\gamma_i$, $c_i$ are given by eqs.(\ref{gammai}),(\ref{luz-i}), respectively, and $\beta_i$ is given by eq.(\ref{ecs_yz1}). If we write down $u$ without the subindex $i$, it means $u=v/c$ for any range of speeds we are dealing with, as for instance in the expressions given by eqs.(\ref{gamma1}),(\ref{gamma2}),(\ref{gammai}) for $\gamma$'s factors. Superluminal regimes have speed intervals of type $c_{i-1}<v\leq c_i$, such that the top (and invariant) speed of the range approximately equals $c_i\simeq\epsilon^{-i+1}c$ -cf. eq.(\ref{luz-i}); therefore, $c_{i+1}/c_i\simeq\epsilon^{-1}$, thus $c_{i+1}\gg c_i$. At speed $c_i$, the top speed of any speed range, the $\gamma_i$ factor takes the value $i\epsilon^{-1}$. We interpret here the states with speed $c_i$ as resonant states of matter, which determine the description of any material body depending on its speed with respect to those of the respective resonant states.\\

Finally, we can say that our spacetime measures given by eqs.(\ref{ecs_xt1}),(\ref{ecs_yz1}) enable us to think of making a discrete transition between the subluminal and the first superluminal regime as stipulated by eq.(\ref{cambio_u2}). To properly carry out it, one needs new dynamical measures compatible with our spacetime measures, so work in this direction should be done if humankind takes this technological challenge as one of its goals.
\bibliographystyle{plainnat}
\bibliography{Artref}
\end{document}